\documentclass[aps,prd,twocolumn,
groupedaddress,amssymb,
showpacs,nofootinbib]{revtex4}

\usepackage{amsmath,amssymb}

\usepackage[usenames]{color}

\usepackage{graphicx}
\usepackage{mathptmx}
\usepackage{bm}
\usepackage{times}

\newcommand{\beq}{\begin{eqnarray}}
\newcommand{\eeq}{\end{eqnarray}}

\begin{document}

\onecolumngrid
\begin{flushright}
BCCUNY-HEP/08-01
\end{flushright}
\twocolumngrid

\title{Near-integrability and confinement
for high-energy hadron-hadron collisions }

%\textcolor{red}{text in red}

%\textcolor{green}{text in green}

\author{Peter \surname{Orland}}

\email{giantswing@gursey.baruch.cuny.edu}

\affiliation{1. The Niels Bohr Institute, The Niels Bohr International Academy,
Blegdamsvej 17, DK-2100, Copenhagen {\O}, Denmark}

\affiliation{2. The Graduate School and University Center, The City University of New York, 365 Fifth Avenue,
New York, NY 10016, U.S.A.}

\affiliation{3. Baruch College, The 
City University of New York, 17 Lexington Avenue, 
New 
York, NY 10010, U.S.A. }

%\date{\today}
\date{January 20, 2008}

\begin{abstract}
We investigate an effective Hamiltonian for QCD at large $s$, in
which longitudinal gauge degrees of freedom are
suppressed, but not eliminated. In an axial gauge the effective field theory
is a set of coupled $(1+1)$-dimensional principal-chiral
models, which are completely integrable. The confinement problem is
solvable in this context, and
we find the longitudinal and transverse string tensions with
techniques already used for a similar Hamiltonian in $(2+1)$-dimensions. We find some
{\em a posteriori} justification
for the effective Hamiltonian as an eikonal approximation. Hadrons
in this approximation consist of partons, which are quarks and
soliton-like excitations of the sigma models. Diffractive 
hadron-hadron scattering appears primarily due to exchange of longitudinal
flux between partons.
\end{abstract}


\maketitle

%\preprint{BCCUNY-HEP/07-06}

%\setcounter{page}{1}

\section{Introduction}
\setcounter{equation}{0}
\renewcommand{\theequation}{1.\arabic{equation}}

In the last decade and a half, effective gauge-theory descriptions
for QCD at large center-of-mass energy squared $s$ has been an active area
of research
\cite{Lipatov}, \cite{McLerranVenugopalan}, \cite{Kovchegov} , \cite{Jalilian-Marian-et-al},
\cite{Jalilian-Marian-et-al-2},
\cite{Verlinde-squared}, 
\cite{KirschLipSzyman}, \cite{Hatta-etal}.
The 
%eikonal 
approximation of Reference \cite{Verlinde-squared} was to eliminate
some gauge-theory degrees of freedom by a longitudinal rescaling. In this paper we
carefully examine the consequences of this rescaling: $x^{0,3}\rightarrow \lambda x^{0,3}$,
$x^{\perp}\rightarrow x^{\perp}$,
$A_{0,3}\rightarrow \lambda^{-1} A_{0,3}$, $A_{\perp}\rightarrow A_{\perp}$, where
$A_{\mu}=A_{\mu}^{a}t_{a}$, $a=1,\dots, N^{2}-1$, denotes the components of the 
SU($N$) Yang-Mills field and the transverse indices
$1,2$ are sometimes collectively denoted by $\perp$. We 
normalize ${\rm Tr}\,t_{a}t_{b}=\delta_{ab}$
and define ${\rm i}f_{ab}^{c}t_{c}=[t_{a},t_{b}]$. The center-of-mass energy squared changes
as $s\rightarrow \lambda^{-2}s$ \cite{Verlinde-squared}. Since 
we wish to consider high energies, we take
$\lambda \ll 1$. 
 
If the scale factor $\lambda$ is small, but
not zero, the resulting Hamiltonian has one extremely small coupling and
one extremely large coupling. The rescaled action is 
\beq
S\!=\!\frac{1}{2g_{0}^{2}}\!\int d^{4}x 
{\rm Tr}\!\left(\! \lambda^{-2}F_{03}^{2}\!+\!\sum_{j=1}^{2}F_{0j}^{2}
\!-\!\sum_{j=1}^{2}F_{j3}^{2} -\lambda^{2} F_{12}^{2}\right)\!,
\label{action}
\eeq
where 
$F_{\mu \nu}=\partial_{\mu}A_{\nu}-\partial_{\nu}A_{\mu}-{\rm i}[A_{\mu},A_{\nu}]$.
The Hamiltonian in $A_{0}=0$ gauge is therefore
\beq
H\!\!&\!\!=\!\!&\!\! \int d^{3} x \left(\frac{g_{0}^{2}}{2}{\mathcal E}_{\perp}^{2}+
\frac{1}{2g_{0}^{2}}{\mathcal B}_{\perp}^{2}+
\frac{(g_{0}^{\prime})^{2}}{2}{\mathcal E}_{3}^{2} \right. \nonumber \\
\!\!&\!\!+\!\!&\!\!
\left. \frac{1}{2(g_{0}^{\prime\prime})^{2}}{\mathcal B}_{3}^{2}\right), \label{ContHamiltonian}
\eeq
where $g_{0}^{\prime}=\lambda g_{0}$, $g_{0}^{\prime\prime}=\lambda^{-1}g_{0}$ 
(so that $g_{0}^{\prime}g_{0}^{\prime\prime}=g_{0}^{2}$),
the electric and magnetic fields are ${\mathcal E}_{i}=-{\rm i}\delta/\delta A_{i}$
and ${\mathcal B}_{i}=\epsilon^{ijk}(\partial_{j}A_{k}+A_{j}\times A_{k})$, respectively
and $(A_{j}\times A_{k})^{a}=f_{bc}^{a}A_{j}^{b}A_{k}^{c}$. Physical states $\Psi$ must 
satisfy Gauss's law
\beq
\left(\partial_{\perp}\cdot {\mathcal E}_{\perp}+\partial_{3}{\mathcal E}_{3}-\rho\right)\Psi=0\;,
\label{Gauss}
\eeq
where $\rho$ is the quark color-charge density. As of this
writing, we are not certain that
this rescaled theory is definitely describing QCD, but think it may 
be a useful phenomenological approach to small-x, large-$s$ scattering.

The Hamiltonian (\ref{ContHamiltonian}) and the constraint
(\ref{Gauss}) will be regularized on a spatial lattice
\cite{Kogut-Susskind}, keeping
time continuous. Afterwards, $x^{3}$  will also be made
continuous, leaving only the transverse coordinates
discrete. We take an axial gauge condition, breaking 
the precedent of considering large-$s$ scattering in
the light-cone gauge. We do this because, as we show later, (\ref{ContHamiltonian}) is
similar to the anisotropic $(2+1)$-dimensional Yang-Mills theory in
this gauge
\cite{PhysRevD71}, \cite{PhysRevD74}, 
\cite{PhysRevD75-1}, \cite{PhysRevD75-2}, 
\cite{CompositeStrings}. If further experience persuades us to use a light-cone lattice
\cite{BardeenPearsonRabinovici} instead, so be it. A light-cone-lattice
approach in this spirit was discussed in References \cite{Aref'eva}. This was 
the starting point of an argument justifying the rescaling
used in (\ref{action}) by an anisotropic renormalization group 
\cite{Aref'evaVolovich}. Similar arguments
were presented for the $(2+1)$-dimensional case \cite{PhysRevD75-2}, 
\cite{CompositeStrings} without knowledge of 
Reference \cite{Aref'evaVolovich}. We should mention that
a lattice gauge theory with different transverse and longitudinal
couplings has also been studied in Reference \cite{Patel}.

The analogous rescaling in $(2+1)$-dimensions gives the 
Hamiltonian, in $A_{0}=0$ gauge\footnote{We note that the coordinate indices
$1$ and $2$ are reversed in References
\cite{PhysRevD71}, \cite{PhysRevD74}, 
\cite{PhysRevD75-1}, \cite{PhysRevD75-2}, 
\cite{CompositeStrings}.}, 
\beq
H\!\!&\!\!=\!\!&\!\! \int d^{2} x \left(\frac{g_{0}^{2}}{2}{\mathcal E}_{1}^{2}+
\frac{1}{2g_{0}^{2}}{\mathcal B}^{2}+
\frac{(g_{0}^{\prime})^{2}}{2}{\mathcal E}_{2}^{2} \right), \label{ContHamiltonian2+1}
\eeq
where $g_{0}^{\prime}=\lambda g_{0}$, and
the electric and magnetic fields are ${\mathcal E}_{i}=-{\rm i}\delta/\delta A_{i}$
and ${\mathcal B}=(\epsilon^{jk}\partial_{j}A_{k}+A_{j}\times A_{j})$, respectively. Gauss's law is
\beq
\left(\partial_{1}\cdot {\mathcal E}_{1}+\partial_{2}{\mathcal E}_{2}-\rho\right)\Psi=0\;.
\label{Gauss2+1}
\eeq

Upon regularization, confinement of quarks can be readily 
demonstrated. Thus the hadronization problem is essentially
solved. This suggests that the elastic part of the
forward amplitude and the soft Pomeron can eventually be understood 
using
our methods.

It has been noted before that the $\lambda \rightarrow 0$ limit of 
QCD is equivalent to a set of principal-chiral ${\rm SU}(N)\!\times \!{\rm SU}(N)$
sigma models  \cite{Verlinde-squared}, \cite{Aref'eva}. Such a sigma model has the 
Lagrangian ${\mathcal L}
=1/(2g_{0}^{2})\eta^{\mu \nu}{\rm Tr} \,\partial_{\mu}U^{\dagger}\partial_{\nu}U$,
where $U\in {\rm SU}(N)$ and $\mu=0,3$. This field theory is integrable, and the
S-matrix \cite{abda-wieg} and even certain off-shell
information (for $N=2$) \cite{KarowskiWeisz} is exactly known. The integrable nature
of large-$s$ scattering is also indicated by the form of Reggeized 
amplitudes
\cite{LipatovKorchemsky}. In Reference \cite{Verlinde-squared} it was pointed
out that the effective gluon-emission vertex \cite{BalitskiFadKurLip} can be derived by
considering longitudinal fluctuations before taking the transverse limit
$\lambda \rightarrow 0$.

The particles of the principal-chiral sigma model are labeled by $r=1,\dots,N-1$
\cite{abda-wieg}.  The particle with label $r$ has an antiparticle 
with label $N-r$. The
mass spectrum is
\beq
m_{r}=m_{1}\frac{\sin\frac{r\pi}{N}}{\sin\frac{\pi}{N}},\;\; m_{1}=K\Lambda(g_{0}^{2}N)^{-1/2}e^{-\frac{4\pi}{g_{0}^{2}N}}+\dots \;, \label{mass-spectrum}
\eeq
where $K$ is a non-universal constant, $\Lambda$ is the ultraviolet cut-off and
the ellipses denote non-universal corrections.

We study the physical states for the effective Hamiltonian. They
are very similar to those of the $(2+1)$-dimensional model
(\ref{ContHamiltonian2+1}), (\ref{Gauss2+1}),
already investigated in detail in
References \cite{PhysRevD71}, \cite{PhysRevD74}, 
\cite{PhysRevD75-1}, \cite{PhysRevD75-2}, 
\cite{CompositeStrings}. Hadrons
consist of quarks joined by electric strings. Longitudinal electric flux consists
essentially of $(1+1)$-dimensional Yang-Mills strings. In contrast, transverse electric
flux is built out of the massive particles of the sigma model. We call these 
Faddeev-Zamolodchikov (FZ) particles, consistent with the terminology
of the operator algebra of the creation and annihilation operators for these
particles. 

In $(2+1)$ dimensions, the transverse and longitudinal string 
tensions\footnote{In References \cite{PhysRevD71}, \cite{PhysRevD74}, 
\cite{PhysRevD75-1}, \cite{PhysRevD75-2}, 
\cite{CompositeStrings}, we called the transverse string
tension the ``vertical string tension" and the longitudinal string tension
the ``horizontal string tension".
} were first found to leading
order in $g_{0}^{\prime}$ \cite{PhysRevD71}. Later, the corrections of
higher order in $g_{0}^{\prime}$
to the longitudinal string tension \cite{PhysRevD74}, and the transverse
string tension \cite{CompositeStrings} were computed. String
tensions for different representations of charges were found and it was shown 
that adjoint sources are not confined
\cite{PhysRevD75-1}. The low-lying mass
spectrum was studied in Reference \cite{PhysRevD75-2}.

Both transverse electric flux
and longitudinal electric flux have some nonzero
thickness. The thickness of
transverse flux is due to color smearing of the FZ particles in the longitudinal
direction \cite{CompositeStrings}. The thickness of longitudinal flux
is caused by creation and destruction of FZ particle pairs, akin to vacuum polarization
\cite{PhysRevD74}. 

After we discuss these confinement mechanisms for the theory
in $(3+1)$ dimensions, we attempt to justify the effective
Hamiltonian as an eikonal approximation. This view was
advocated in Reference 
\cite{Verlinde-squared}, but we interpret the meaning of the
factor $\lambda$ slightly differently.

Each of the two incoming hadrons in a collision process
is a collection of FZ particles, joined by longitudinal-electric 
flux lines. Scattering of two hadrons can involve rearrangement of the
flux lines among the FZ particles. It is also
possible for a non-Abelian
phase rotation of FZ particles to occur as they pass through each other; the phase
factor is given by the exact $(1+1)$-dimensional exact S-matrix \cite{abda-wieg}. These
two processes are not entirely distinct, as can be seen from an examination of the
$1/N$ expansion of this S-matrix. 
%Scattering
%of large nuclei will produce many processes of the second type \cite{McLerranVenugopalan},
%resulting in a color-glass-condensate picture \cite{ColorGlass}. 

In the next section, we put the Hamiltonian (\ref{ContHamiltonian}) on a spatial
lattice, then transform to the axial gauge. In Section 3, we 
take the continuum limit in the $x^{3}$-direction and show that
this Hamiltonian is equivalent to a collection of integrable field theories which
are coupled
together. We discuss confinement of
color and find the transverse string tension in Section 4. We find
the longitudinal string
tension in Section 5. By comparing 
the electric field strengths in different directions, we find some justification of
the effective Hamiltonian in Section 6. In Section 7, we describe the nature of
hadronic states and begin an 
examination of diffractive hadronic scattering, focussing on the exchange
of hadronic flux. We present some conclusions and mention some
further directions for research in Section 8.

\section{Regularization and the axial gauge}
\setcounter{equation}{0}
\renewcommand{\theequation}{2.\arabic{equation}}

Consider a lattice of sites $x$, whose coordinates are $x^{j}$, $j=1,2,3$, where
$x^{j}/a$
are integers, and where
$a$ is the lattice spacing. Each link is a pair $x$, $j$, and joins the site 
$x$ to $x+{\hat j}a$, where
$\hat j$ is a unit vector in the $j^{\rm th}$ direction. We choose
temporal gauge $A_{0}=0$. Before 
any spatial gauge fixing, the degrees of freedom are elements of the group SU($N$) in the 
fundamental ($N\times N$)-dimensional 
matrix representation $U_{j}(x)\in$ SU($N$) at each link
$x$, $j$. In addition, there are the electric-field 
operators at each link
$l_{j}(x)_{b}$, $b=1,\dots, {N}^{2}-1$. The 
commutation relations on the
lattice are
\begin{eqnarray}
[l_{j}(x)_{b} , l_{k}(y)_{c} ]=
i\delta_{x\;y}\delta_{j\;k} \;f_{b c}^{d}
\;l_{j}(x)_{d} \;, \nonumber 
\end{eqnarray}
\begin{eqnarray}
[l_{j}(x)_{b}, U_{k}(y)]        =
-\delta_{x\;y}\delta_{j\;k}\; t_{b}\;U_{j}(x)\;,
\label{loccommrel}
\end{eqnarray}
all others zero.

The lattice version of (\ref{ContHamiltonian})  is $H=H_{0}+H^{\prime}+H^{\prime\prime}$, where
\begin{eqnarray}
H_{0}\!\!&\!\!=\!\!& \!\!\frac{1}{a}\sum_{x} 
\left[ \frac{g_{0}^{2}}{2} l_{\perp}(x)^{2}-
\sum_{j=1,2}\frac{1}{4g_{0}^{2}}\; {\rm Tr}\; U^{\square}_{j}(x) 
\right] \;,\nonumber \\
H^{\prime}\!\!&\!\!=\!\!&\!\! \frac{(g_{0}^{\prime})^{2}}{2a}\sum_{x} l_{3}(x)^{2},\;
H^{\prime\prime}=-\frac{1}{4(g_{0}^{\prime\prime})^{2}a}\sum_{x} {\rm Tr}\; U^{\square}_{3}(x) 
, \label{hamilt1}
\end{eqnarray}
and where, as before, 
$g_{0}^{\prime}=\lambda g_{0}$, $g_{0}^{\prime\prime}=\lambda^{-1}g_{0}$  and
\begin{eqnarray}
U^{\square}_{i}(x)  =
\epsilon^{ijk}U_{j}(x)
U_{k}(x+{\hat j}a)
U_{j}(x+{\hat k}a)^{\dagger}
U_{k}(x)^{\dagger} \;.
\label{plaquette}
\end{eqnarray}

We need the adjoint representation of the SU($N$) gauge field ${\mathcal R}_{j}(x)$, defined by
${\mathcal R}_{j}(x)_{b}^{\;\;c}t_{c}=
U_{j}(x)t_{b}U_{j}(x)^{\dagger}$. This has the properties
${\mathcal R}_{j}(x)\in {\rm SU}(N)/{\mathbb Z}_{N}$, 
${\mathcal R}_{j}(x)^{\rm T}{\mathcal R}_{j}(x)=1$, 
and ${\rm det}\;{\cal R}_{j}(x)=1$. Notice that
\begin{eqnarray}
[{\mathcal R}_{j}(x)_{b}^{\;\;c} l_{j}(x)_{c}, U_{k}(y)]        =
-\delta_{x\;y}\delta_{j\;k}\; U_{j}(x)\;t_{b}\;.
\nonumber
\end{eqnarray}
Thus $l_{j}(x)$ generates infinitesimal SU($N$) transformations on the
left of $U_{j}(x)$ and ${\mathcal R}_{j}(x)l_{j}(x)$ generates infinitesimal 
SU($N$) transformations on the
right of $U_{j}(x)$. The squares of these operators are the same
\beq
[l_{j}(x)]^{2}=[{\mathcal R}_{j}(x)l_{j}(x)]^{2}\;, \nonumber
\eeq
by virtue of the orthogonality of the adjoint representation.

Color charge operators, denoted by $q(x)_{b}$, satisfy
\begin{eqnarray}
[q(x)_{b},q(y)_{c}]={\rm i} f_{bc}^{d} \delta_{xy}q(x)_{d} \; . \label{charge-comm}
\end{eqnarray}

Schr\"odinger wave functions are  
complex-valued functions
of {\em all} the link degrees of freedom
$U_{j}(x)$. Physical wave functions $\Psi(\{U\})$ satisfy Gauss' law
\begin{eqnarray}
\left[( {\cal D} \cdot l)(x)_{b}-q(x)_{b}\right]\; \Psi( \{U \}) =0 \;.
\label{gauss}
\end{eqnarray}
where (with no summation of $j$)
\begin{eqnarray}
\left[{\cal D}_{j} l_{j}(x)\right]_{b}  = 
l_{j}(x)_{b}-
{\cal R}_{j}(x-{\hat j}a)_{b}^{\;\;\;c} \; l_{j}(x-{\hat j}a)_{c} \;. \label{cov-deriv}
\end{eqnarray}

Next we impose the axial gauge condition $U_{3}=1$. We take the lattice to be open at
$x^{3}=0, L^{3}$, which means we do not fix any non-contractible Wilson loops. The
open boundary condition means, however, that a relic of Gauss's law must still be
imposed.

We
choose space to be a lattice ``cylinder" of size 
$L^{1}\times L^{2}\times L^{3}$, with periodic boundary
conditions in the $1$- and $2$-directions only. This means that for any function of
position
$f(x)$, we have $f(x^{1}+mL^{1},x^{2}+nL^{2},x^{3})=f(x)$, for
any $m,n\in {\mathbb Z}$. We take
components of $x$ to have the values 
$x^{1,2}=0,a,2a,\dots ,L^{1,2}-a$, and $x^{3}=0,a,2a,\dots, L^{3}$. Gauss's 
law is still given by (\ref{gauss}), provided
(\ref{cov-deriv}) is modified to
\begin{eqnarray}
{\cal D}_{3} l_{3}(x) \!\!&\!\!\! =\!\!\!&\!\!  \delta_{x^{3}\neq L^{3}}l_{3}(x)
-\delta_{x^{3}\neq 0}{\cal R}_{3}(x^{\perp},x^{3}\!-\!a) l_{3}(x^{\perp},x^{3}\!-\!a),
\nonumber \\
{\cal D}_{1} l_{1}(x) \!\!&\!\! =\!\!&\!\! l_{1}(x)
- {\cal R}_{1}(x^{1}-a,x^{2},x^{3}) l_{1}(x^{1}-a,x^{2},x^{3})\;, 
\nonumber \\
{\cal D}_{2} l_{2}(x) \!\!&\!\! =\!\!&\!\! l_{2}(x)
-  {\cal R}_{2}(x^{1},x^{2}-a,x^{3})l_{2}(x^{1},x^{2}-a,x^{3})\;.
\label{cov-deriv1}
\end{eqnarray}

To fix the links in the $3$-direction, we need to use
(\ref{gauss}) and (\ref{cov-deriv1}) to solve for $l_{3}$:  
\begin{eqnarray}
l_{3}(x) =  
\sum_{y^{3}=0}^{x^{3}}\left[ q(x^{\perp},y^{3})- ({\cal D}_{\perp} \cdot l_{\perp})
(x^{\perp}, y^{3})\right]
\;. \label{cylgausssoln}
\end{eqnarray}

Some non-Abelian gauge invariance remains, namely
\begin{eqnarray}
\Gamma(x^{\perp}) \Psi= 
\sum_{x^{3}=0}^{L^{3}}
\left[( {\cal D}_{\perp} \cdot l_{\perp})(x)-q(x)\right]\Psi=0
\;. \label{remaining}
\end{eqnarray}
This remaining gauge invariance means that (\ref{cylgausssoln}) is equivalent
to
\begin{eqnarray}
l_{3}(x) =  
-\sum_{z^{3}=x^{3}+a}^{L^{3}}\left[ q(x^{\perp},z^{3})- ({\cal D}_{\perp} \cdot l_{\perp})
(x^{\perp}, z^{3})\right]
\;. \label{cylgausssoln1}
\end{eqnarray}

The new expressions for $l_{3}$, namely (\ref{cylgausssoln}) and
(\ref{cylgausssoln1}), allow us to completely
eliminate all degrees of freedom but $U_{2,3}$. We now can write the term in
the Hamiltonian (\ref{hamilt1}) which depends on the longitudinal electric field as
\beq
H^{\prime}=
\!\!&\!\!-\!\!&\!\!
\frac{(g_{0}^{\prime})^{2}}{2a}\sum_{x^{\perp}}
\sum_{y^{3}=0}^{x^{3}} \,\sum_{z^{3}=x^{3}+a}^{L^{3}}
\left[ q(x^{\perp},y^{3})- ({\cal D}_{\perp} \cdot l_{\perp})
(x^{\perp}, y^{3})\right] 
\nonumber \\
\!\!&\!\!\times \!\!&\!\! 
\left[ q(x^{\perp},z^{3})-( {\cal D}_{\perp} \cdot l_{\perp})
(x^{\perp}, z^{3})\right] \;, \nonumber
\eeq
or
\begin{widetext}
\beq
H^{\prime}
=-\frac{(g_{0}^{\prime})^{2}}{4a^{2}}\sum_{x^{\perp}}\;
\sum_{y^{3}, z^{3}=0}^{L^{3}}\; \vert y^{3}-z^{3} \vert
\left[ q(x^{\perp},y^{3})- ({\cal D}_{\perp} \cdot l_{\perp})
(x^{\perp}, y^{3})\right]  
\left[ q(x^{\perp},z^{3})- ({\cal D}_{\perp} \cdot l_{\perp})
(x^{\perp}, z^{3})\right] \;, \label{LonElectric}
\eeq
\end{widetext}
In the axial gauge, the longitudinal-electric-field-squared
term $H^{\prime}$ is highly non-local in the $3$-direction. This non-locality is
a standard feature of physical gauges. This fact has a simple physical
interpretation in
the case of (\ref{LonElectric}). The longitudinal electric field has been
eliminated in favor of the transverse degrees of freedom. Now electric flux between
two charged transverse plates  must be proportional to the longitudinal
separation
of these plates. This is accounted for by
the linear factor in (\ref{LonElectric}). On the other hand, 
the vacuum expectation value of $H^{\prime}$ must be proportional to the
spacial volume $L^{1}L^{2}L^{3}$. The non-locality of (\ref{LonElectric})
means that the nature of the vacuum state is subtle. Mandelstam, who considered
the analogous continuum Hamiltonian,  
argued that the vacuum state can only have finite energy
density if magnetic condensation takes place
\cite{Mandelstam}. His reasoning was that this is the way in which the vacuum
correlator 
\beq
\left<  0 \right\vert [ q(x^{\perp},y^{3})\!\!\!&\!\!\!-\!\!\!&\!\!\! ({\cal D}_{\perp} \cdot l_{\perp})
(x^{\perp}, y^{3}) ]    \nonumber \\
&\times& [ q(x^{\perp},z^{3})- ({\cal D}_{\perp} \cdot l_{\perp})
(x^{\perp}, z^{3}) ]
\left\vert 0 \right>\;, \nonumber
\eeq
can fall off sufficiently quickly in $\vert y^{3}-z^{3} \vert$. 

What of the remainder
of the Hamiltonian $H_{0}+H^{\prime\prime}$ after gauge fixing? Setting $U_{1}(x)=1$, we
find
\beq
H_{0}=\sum_{x^{\perp}} \left[ H_{0}(x^{\perp},2)+H_{0}(x^{\perp},3) \right]\;,
\label{transverseH1}
\eeq
where
\beq
H_{0}(x^{\perp}, \!\!\!\!&\!\!j\!\!&\!\!\!\!
)    
=
\sum_{x^{3}=0}^{L^{3}}\frac{g_{0}^{2}}{2a} l_{j}(x)^{2} \nonumber \\
\!\!&\!\!-\!\!&\!\!
\sum_{x^{3}=0}^{L^{3}-a}\frac{1}{2g_{0}^{2}a}\; {\rm Re\;Tr}\; U_{j}(x) 
U_{j}(x^{\perp}, x^{3}+a)^{\dagger} .
\label{transverseH2}
\eeq
On the other hand, the longitudinal magnetic
term $H^{\prime\prime}$, is unchanged, since it
depends only on the transverse components
of the gauge field.

Now that the Hamiltonian has been recast in an axial gauge, we may, at least
in principle, take the thermodynamic limit, in which 
$L^{1}$, $L^{2}$ and $L^{3}$ all become infinity.

\section{Coupled $(1+1)$-dimensional field theories}
\setcounter{equation}{0}
\renewcommand{\theequation}{3.\arabic{equation}}

Next let us examine each of the terms of the Hamiltonian more closely. We will
take a continuum limit in the ${3}$-direction. Thus only the transverse coordinates
will be latticized. The resulting structure resembles a transit box for bottles of wine
as
shown in Fig. 1. Each 
term $H_{0}(x^{\perp},j)$, 
defined in  (\ref{transverseH2}), is a lattice
$(1+1)$ principal-chiral nonlinear sigma model, with
coupling constant $g_{0}$. We
will regard $H_{0}$ as the Hamiltonian of the unperturbed theory and treat
$H^{\prime}$ and $H^{\prime\prime}$ as interactions. Notice that
the coefficients of each of these terms is small, by virtue of $\lambda \ll1$. Thus one
coupling, $g_{0}$ is small (to take the continuum limit), another coupling $g_{0}^{\prime}$
is much smaller still and the third coupling $g_{0}^{\prime\prime}$ is
comparatively extremely large. We shall say more about the sizes of the
couplings in the next section.

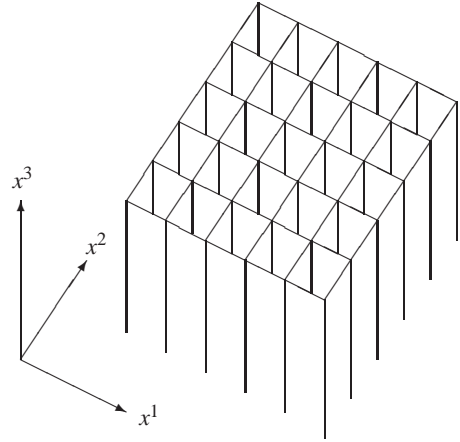
\begin{figure}[ht]

\begin{picture}(150,200)(45,0)

\put(10,40){\vector(2,-1){40}}
\put(55,15){$x^{1}$}
\put(10,40){\vector(2,3){25}}
\put(35,80){$x^{2}$}
\put(10,40){\vector(0,1){60}}
\put(7,105){$x^{3}$}

%\linethickness{0.5mm}

\multiput(50,100)(10,15){6}{\line(2,-1){75}}
\multiput(50,100)(15,-7.5){6}{\line(2,3){50}}

\multiput(0,50)(15,-7.5){6}{\multiput(50,50)(10,15){6}{\line(0,-1){20}}}

\multiput(50,100)(15,-7.5){6}{\line(0,-1){50}}

\multiput(125,60)(10,15){6}{\line(0,-1){50}}

\end{picture}

\caption{Space with $x^{\perp}$ discrete and $x^{3}$ continuous
resembles the interior of a wine transit box. }
\end{figure}

%Some simple intuitive reasoning yields the leading form of the 
%potential between a quark and antiquark
%for anisotropic gauge theories in $(2+1)$ dimensions
%\cite{PhysRevD71}. For this dimensionality, all
%couplings are weak. String tensions for higher
%representations can also be worked out, and it is not hard to see that
%adjoint sources are
%not confined  \cite{PhysRevD75-1}. The string tensions have corrections, however, which
%can be explicitly computed for
%gauge group SU($2$). The 
%leading
%correction to the longitudinal 
%string tension was found in Reference \cite{PhysRevD74}. The leading correction to the 
%transverse string tension was found very recently \cite{CompositeStrings}. The
%low-lying glueball spectrum can also be determined \cite{PhysRevD75-2}. Much
%of what we shall do in this and the next
%two sections is extend this earlier work to the $(3+1)$-dimensional
%case.

We will assume the lattice spacing is small and treat the principal-chiral sigma
models as near their thermodynamic and 
continuum limits. Each sigma model lives in a {\em band}, that is a
a two-dimensional strip of length $L^{3}\rightarrow \infty$ and width $a$, in the
$j$-$3$ plane, where $j=1,2$ (we referred to bands as layers in the papers 
on $(2+1)$-dimensional gauge theories. In three spatial dimensions, a different
name seems appropriate). In our wine-transit-box analogy, there are four bands
surrounding each wine bottle, and two wine bottles adjacent to each band. We 
denote the band between the line at $x^{\perp}$
and the line at $x^{\perp}+{\hat j}a$ by $(x^{\perp},j)$. The left-handed 
and right-handed currents of sigma model in the band $(x^{\perp},j)$ are 
\beq
{\mathcal J}^{\rm L}_{\mu}(x^{\perp}, x^{3},j)_{b}&=&
{\rm i}\,{\rm Tr}\,\,t_{b} \, \partial_{\mu}U_{j}(x^{\perp},x^{3})\, 
U_{j}(x^{\perp},x^{3})^{\dagger}\;, \nonumber \\
{\mathcal J}^{\rm R}_{\mu}(x^{\perp}, x^{3},j)_{b}&=&{\rm i}\,{\rm Tr}\,\,t_{b} \, 
U_{j}(x^{\perp}, x^{3})^{\dagger}\partial_{\mu}U_{j}(x^{\perp}, x^{3})\;,
\nonumber 
\eeq 
respectively, 
where $\mu=0,3$. The operator ${\mathcal J}^{\rm L}_{0}(x^{\perp}, x^{3},j)$
produces an SU($N$) rotation at the edge of the band $(x^{\perp},j)$ at
$x^{\perp}$. The operator ${\mathcal J}^{\rm R}_{0}(x^{\perp}, x^{3},j)$
produces an SU($N$) rotation at the other edge of the band $(x^{\perp},j)$ at
$x^{\perp}+{\hat j}a$.

We write $H_{0}$ as
\beq
H_{0}
=
%\!\!&\!\!=\!\!&\!\! 
\sum_{x^{\perp}, j}\int dx^{3} 
\frac{1}{2g_{0}^{2}}[ {\mathcal J}^{\rm L}_{0}(x^{\perp}, x^{3},j)^{2}
%\nonumber \\
%\!\!&\!\!+\!\!&\!\! 
\!\!+\!\!
{\mathcal J}^{\rm L}_{3}(x^{\perp}, x^{3},j)^{2}]. \label{HNLSM}
\eeq
The connection between (\ref{transverseH1}), (\ref{transverseH2}) and
(\ref{HNLSM}) is made through the Heisenberg equation of motion
for $U_{j}$. This gives 
$l_{j}(x)\approx ag_{0}^{-2}{\mathcal J}^{\rm L}_{0}(x^{\perp}, x^{3},j)$, for small $a$. Similarly,
${\mathcal R}_{j}(x)l_{j}(x)\approx ag_{0}^{-2}{\mathcal J}^{\rm R}_{0}(x^{\perp}, x^{3},j)$. The 
transverse
electric field is now ${\mathcal J}^{LR}_{0}$.

The  $\;$ residual gauge-invariance  condition
and
the  longitudi-\\nal-electric-field-squared term are written by the same substitution
into (\ref{remaining}) and (\ref{LonElectric}), respectively. Residual gauge invariance is
now
the condition on physical states $\Psi$ 
\beq
\int dx^{3}\sum_{j=1,2}
\left[ {\mathcal J}^{\rm L}_{0}(x^{\perp},x^{3},j)\right.\!\!\!&\!\!\!-\!\!\!&\!\!\! 
{\mathcal J}^{\rm R}_{0}(x^{\perp}-{\hat j}a,x^{3},j) \nonumber \\
\!\!\!&\!\!\!-\!\!\!&\!\!\!\left. \rho(x^{\perp},x^{3})
\right]\!\Psi\!=\!0, \label{continuumConstraint}
\eeq
and the longitudinal-electric-field-squared term is
\begin{widetext}
\beq
H^{\prime}=
&-&\frac{(g_{0}^{\prime})^{2}}{4g_{0}^{4}a^{2}}\sum_{x^{\perp}}\sum_{j=1,2}  \int \! dx^{3}\!\!\int \!dy^{3}
\, \vert x^{3}-y^{3}\vert 
\left[ {\mathcal J}^{\rm L}_{0}(x^{\perp},x^{3},j)- {\mathcal J}^{\rm R}_{0}(x^{\perp}-{\hat j}a,x^{3},j)
-\rho(x^{\perp},x^{3})
\right] \nonumber \\
&\times&\left[ {\mathcal J}^{\rm L}_{0}(x^{\perp},y^{3},j)- 
{\mathcal J}^{\rm R}_{0}(x^{\perp}-{\hat j}a,y^{3},j)
-\rho(x^{\perp},y^{3})
\right] 
\label{continuumLonElectric}
\eeq
\end{widetext}
where $\rho(x^{\perp}, x^{3})$ is a linear charge density, satisfying the
algebra
\beq
[\rho(x^{\perp}, x^{3})_{b},\rho(y^{\perp}, y^{3})_{c}]
&=&i\delta_{x^{\perp} y^{\perp}}\delta(x^{3}-y^{3})\nonumber \\
&\times&f^{d}_{bc}\rho(x^{\perp}, x^{3})_{d} \;.
\label{charge-density-algebra}
\eeq
The last term in the Hamiltonian is the longitudinal-magnetic-field-squared term
\beq
H^{\prime\prime}=-\frac{1}{4(g_{0}^{\prime\prime})^{2}a^{2}}\sum_{x^{\perp}}
\int dx^{3} \;{\rm Re\;Tr}\; U^{\square}(x^{\perp},x^{3})
\;,
\label{continuumLonMagnetic}
\eeq
where $ U^{\square}(x^{\perp},x^{3})$ is given by (\ref{plaquette}), as before.
%\beq
%U^{\square}(x^{\perp},x^{3})=U_{1}(x^{\perp},x^{3})U_{2}(x^{\perp}+{\hat 1}a,x^{3})
%U_{1}(x^{\perp}+{\hat 2}a,x^{3})^{\dagger}U_{2}(x^{\perp},x^{3})^{\dagger}.
%\nonumber
%\eeq

\begin{figure}[ht]

\begin{picture}(150,200)(50,50)

%\linethickness{0.5mm}

\put(55,90){\vector(2,-1){40}}
\put(97,65){$x^{1}$}
\put(55,90){\vector(2,3){25}}
\put(80,130){$x^{2}$}
\put(55,90){\vector(0,1){60}}
\put(53,155){$x^{3}$}

\put(100,100){\line(2,-1){10}}

\put(130,85){\line(2,-1){30}}

\put(110,55){\line(2,3){20}}

\put(110,155){\line(2,3){40}}

\put(100,200){\line(2,-1){60}}

\put(100,100){\line(0,1){100}}

\put(130,85){\line(0,1){100}}

\put(160,70){\line(0,1){100}}

\put(110,55){\line(0,1){100}}

\put(150,175){\line(0,1){40}}

\end{picture}

\caption{The four bands meeting at the line of fixed $x^{\perp}$. }

\end{figure}
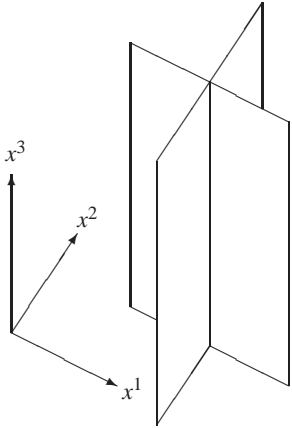

The terms in the Hamiltonian (\ref{HNLSM}), 
(\ref{continuumLonElectric}), (\ref{continuumLonMagnetic}) and the constraint 
(\ref{continuumConstraint}) may seem a bit complicated, but they each have a straightforward
geometrical interpretation. A sigma model lives in each band $(x^{\perp},j)$, whose Hamiltonian
is $H_{0}(x^{\perp}, j)$. The 
excitations of $H_{0}$ are the FZ particles, which behave like 
solitons, though they are not quantized versions of classical solutions. Without the
interaction terms, these particles move only in the $3$-direction, scattering 
with a nontrivial
S-matrix. The scattering is integrable, which implies that
there is no particle creation or destruction. The elementary
$r=1$ FZ particles are adjoint gluon-like
particles (see equation (\ref{mass-spectrum})). They 
can be thought of a color dipoles with a right fundamental color
charge (anti-charge) at $x^{\perp}$ and a left fundamental anti-charge (charge)
at $x^{\perp}+{\hat j}a$. All other FZ particles can be built out of these $r=1$ ``diffractive gluons".

We can regard a line in the $3$-direction as
a choice of $x^{\perp}$. Consider now the four bands which
meet at this line $x^{\perp}$, namely $(x^{\perp},1)$, $(x^{\perp},2)$,
$(x^{\perp}-{\hat 1}a,1)$ and $(x^{\perp}-{\hat 2}a,2)$, shown in Fig. 2. There are five
color charges at the line $x^{\perp}$, one from each band and
one from the color source $\int dx^{3} \rho(x^{\perp},x^{3})$ on the line. The 
nontrivial constraint (\ref{continuumConstraint}) means that sum of these
five color charges is zero. 

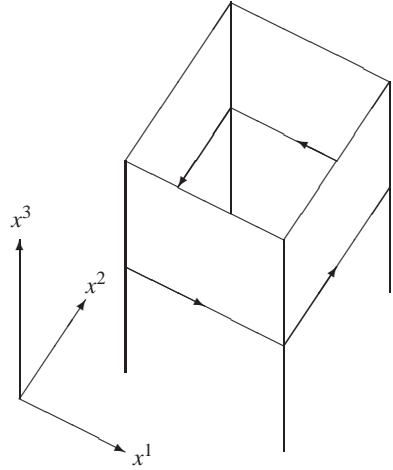
\begin{figure}[hb]

\begin{picture}(150,200)(50,50)

%\linethickness{0.5mm}

\put(40,90){\vector(2,-1){40}}
\put(83,65){$x^{1}$}
\put(40,90){\vector(2,3){25}}
\put(65,130){$x^{2}$}
\put(40,90){\vector(0,1){60}}
\put(37,155){$x^{3}$}

\put(80,180){\line(2,-1){60}}
\put(80,180){\line(2,3){40}}
\put(120,240){\line(2,-1){60}}
\put(140,150){\line(2,3){40}}

\put(120,160){\line(0,1){80}}
\put(80,100){\line(0,1){80}}

\put(140,70){\line(0,1){80}}

\put(180,130){\line(0,1){80}}

\linethickness{1.5mm}
\put(80,140){\line(2,-1){60}}
\put(80,140){\vector(2,-1){30}}
\put(100,170){\line(2,3){20}}
\put(120,200){\vector(-2,-3){20}}
%\put(120,200){\line(2,-1){60}}
\put(140,110){\line(2,3){40}}
\put(140,110){\vector(2,3){20}}
\put(160,180){\vector(-2,1){15}}
\put(160,180){\line(-2,1){40}}

\end{picture}

\caption{The Wilson loop $U^{\square}(x^{\perp},x^{3})$, directed
counter-clockwise around four bands.}

\end{figure}

The interaction $H^{\prime}$, as discussed before, is simply the 
${\mathcal E}_{1}^{2}$ term of the Hamiltonian in axial gauge.

FInally, the interaction $H^{\prime\prime}$ is a discrete version of the integral of the
square of the longitudinal
magnetic flux. The quantity ${\rm Tr} U^{\square}(x^{\perp},x^{3})$ is the
Wilson loop around four bands (a wine bottle stands in the middle of these
four bands), shown in Fig. 3.

\section{Transverse confinement}
\setcounter{equation}{0}
\renewcommand{\theequation}{4.\arabic{equation}}

Next we explain how transverse confinement works for our effective
gauge theory. The mechanism does not rely on our taking the continuum
limit of the coordinate $x^{3}$; it holds just as well on the three-dimensional
lattice. The key ingredients are the mass gap of $H_{0}$ and
residual gauge invariance (\ref{remaining}) or (\ref{continuumConstraint}). We will
show that our explanation makes sense for
\beq
(g_{0}^{\prime})^{2} =\frac{g_{0}^{4}}{(g_{0}^{\prime\prime})^{2}}\ll \frac{1}{g_{0}}e^{-4\pi/(g_{0}^{2}N)}\;. \label{relative-scales}
\eeq
Our arguments are adapted from Reference \cite{PhysRevD71}.

Let us first consider the extreme case of $\lambda=0$. In this case, $g_{0}^{\prime}=0$
and $g_{0}^{\prime\prime}=\infty$, so that $H=H_{0}$. Let us place a quark at
$u^{\perp}, u^{3}$ and an antiquark at $v^{\perp},v^{3}$. We now ask what
the ground-state energy is, with these sources introduced. This energy is an
eigenvalue of $H_{0}$. Since $H_{0}$ is the sum of nonlinear sigma models, we try to put
as many as possible of these sigma models in their ground states. In other words,
we want as many bands of our wine transit box 
as possible to be unoccupied by FZ particles. Consider the effect of the residual
condition on states, (\ref{continuumConstraint}). This tells us
that we cannot make all four of the bands meeting at $u^{\perp}$ in a color-singlet
state, due to the presence of the quark at $u^{\perp}$. Now we want the sigma model
in each of these
bands $(u^{\perp},1)$, $(u^{\perp},2)$, $(u^{\perp}-{\hat 1}a,1)$ and $(u^{\perp}-{\hat 2}a,2)$
to be in an eigenstate (since we seek the lowest-energy eigenstate of $H_{0}$). Now
at least one of these four sigma models cannot be in a color-singlet state. The 
Hohenberg-Mermin-Wagner theorem tells us that there is no spontaneous
symmetry breaking for the vacuum of a $(1+1)$-dimensional field theory. Hence the
vacuum must be a singlet. Any non-singlet eigenstate must therefore have 
energy of at least the mass gap, $m_{1}$. So at least one of the four 
sigma models must contain an FZ particle, say $(u^{\perp},1)$. If there is no
color source in the line $u^{\perp}+{\hat 1}a$, then by the constraint (\ref{relative-scales}),
at least one of the other three sigma models in bands adjacent to this line,
namely 
$(u^{\perp}+{\hat 1}a,1)$, $(u^{\perp}+{\hat 1}a,2)$ and $(u^{\perp}+{\hat 1}a-{\hat 2}a,2)$,
must also be excited, by the same reasoning. Continuing in the this way, there
must be a connected two-dimensional union of bands in which all the sigma models
are excited, terminating at $u^{\perp}$. For the energy to be finite, this
two-dimensional union of bands must also terminate at $v^{\perp}$, where the antiquark is
present. 

The physical picture of transverse confinement is the following. Imagine we
look at the quark-antiquark pair from a great distance along the $3$-direction. We see a string
of FZ particles in the transverse plane, joining the quark to the antiquark, as in Fig. 4. The 
potential
is the sum of the masses of all these FZ particles. As it happens, the potential is
not rotation invariant, even in the transverse plane. If the quark and antiquark
are separated along the $1$- or $2$-directions, however, 
the potential between these sources is linear with transverse
string tension
\beq
\sigma_{\rm \perp}=\frac{m_{1}}{a}\;. \label{transversesigma}
\eeq
The 
lack of rotational invariance around the $x^{3}$-axis seems unrealistic, but
we will argue at the end of
this section that this invariance comes about when $H^{\prime}$ and $H^{\prime\prime}$ are
included.

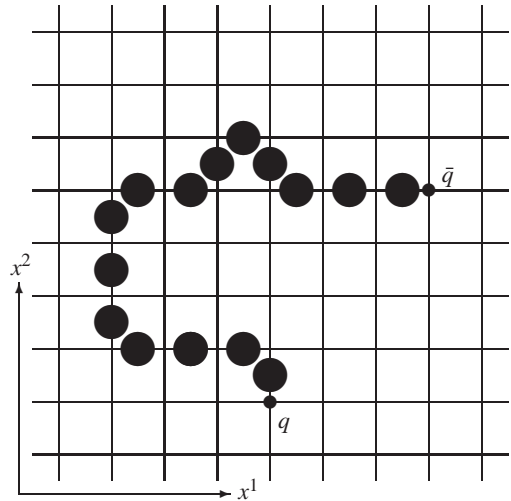
\begin{figure}[ht]

\begin{picture}(150,200)(30,0)

\multiput(10,20)(0,20){9}{\line(1,0){180}}

\multiput(20,10)(20,0){9}{\line(0,1){180}}

\put(5,5){\vector(1,0){80}}
\put(88,3){$x^{1}$}

\put(5,5){\vector(0,1){80}}
\put(2,88){$x^{2}$}

\put(103,30){$q$}

\put(165, 123){$\bar q$}

\put(100,40){\circle*{5}}
\put(100,50){\circle*{13}}
\put(90,60){\circle*{13}}
\put(70,60){\circle*{13}}
\put(70,60){\circle*{13}}
\put(50,60){\circle*{13}}
\put(40,70){\circle*{13}}
\put(40,90){\circle*{13}}
\put(40,110){\circle*{13}}
\put(50,120){\circle*{13}}
\put(70,120){\circle*{13}}
\put(100,130){\circle*{13}}
\put(90,140){\circle*{13}}
\put(80,130){\circle*{13}}
\put(110,120){\circle*{13}}
\put(130,120){\circle*{13}}
\put(150,120){\circle*{13}}
\put(160,120){\circle*{5}}

\end{picture}
\caption{The transverse projection of a transverse string, built out of
FZ particles (represented by large bullets), joining a quark to an antiquark
(represented by small bullets). Longitudinal flux lines
join the particles together. These flux lines are parallel to the
line of sight, hence are not visible. The neighboring FZ particles and sources
in this figure are not adjacent, in general,
since they may be located at different values of $x^{3}$.}

\end{figure}

We can readily see that certain spacelike Wilson loops have an area law. Consider
a rectangular loop in the $x^{1}$-$x^{3}$-plane. By virtue of the gauge condition
$U_{3}=1$, this loop breaks apart into a product of sigma-model correlation functions. Suppose
the dimensions of the loop are $M^{1}$ and $M^{3}$ in the $1$- and $3$-directions, respectively.
Then the Wilson loop is roughly (in that we are being sloppy with contractions of
group indices)
\beq
A \sim \prod_{x^{3}=y^{3}}^{y^{3}+M^{3}}
\left< 0\right\vert U_{1}(x^{\perp}, x^{3})U_{1}(x^{\perp}+{\hat 1}M^{1},x^{3})^{\dagger} \left\vert 0 \right>\;.
\nonumber
\eeq
Each of the correlation functions in this product decays exponentially as
\beq
\left< 0\right\vert U_{1}(x^{\perp}, x^{3})U_{1}(x^{\perp}+{\hat 1}M^{1},x^{3})^{\dagger} \left\vert 0 \right>
\sim e^{-m_{1}M^{1}}\;, \nonumber 
\eeq
where $\left\vert 0 \right>$ is the vacuum of the sigma model. Therefore 
the Wilson loop behaves as
\beq
A \sim \exp{-\sigma_{\perp}M^{1}M^{3}} \;.\label{AreaLaw}
\eeq
Loops oriented in the $x^{2}$-$x^{3}$-plane behave exactly the same way. 

Let us now suppose we increase $\lambda$ a bit, so that it is no longer
zero, but so that (\ref{relative-scales}) is satisfied. Then the coefficients of
$H^{\prime}$ and $H^{\prime\prime}$  are small compared to $m_{1}/a$. We
have three mass scales in the problem, namely $m_{1}$, $(g_{0}^{\prime})^{2}/a$
and $1/[(g_{0}^{\prime\prime})^{2} a]$. If the first of these mass scales is much greater
than the other two, transverse confinement will still hold. To determine the 
correction to the potential, as $g_{0}^{\prime}$ is increased and $g_{0}^{\prime\prime}$
is decreased will require the application of exact matrix elements; current
form factors for $H^{\prime}$ \cite{KarowskiWeisz}
and field form factors for $H^{\prime\prime}$ \cite{BalogWeisz}. The inclusion of
these terms will mean that FZ particles interact between adjacent bands. There
are also be processes whereby FZ particles can be created or destroyed, due to
these interactions. These effects mean that the theory is no longer integrable, though
it is nearly so, provided
(\ref{relative-scales}) holds. 

The correction to the transverse string tension for $(2+1)$-dimensional
SU($2$) gauge theory comes from
the lower-dimensional analogue of $H^{\prime}$ (there is no term analogous to
$H^{\prime\prime}$) \cite{CompositeStrings}. This correction is due to
fluctuations of positions
of FZ particles, bound by longitudinal strings. The longitudinal-string potential 
(which we discuss in the next section) is
not linear at short distances, but quadratic; this is due to fact
that the color of an FZ particle is smeared in the longitudinal direction. The
actual distribution of color is approximately Gaussian, which can be found
using a form factor for the current operator of the sigma model \cite{KarowskiWeisz}. 

We expect
that corrections coming from the inclusion
of $H^{\prime\prime}$
make the potential invariant with respect to rotations about the
$3$-axis. This is because $H^{\prime\prime}$ acting on string states
deforms contour of the string in the transverse plane. Perturbation
theory in this term should therefore cause
the string to fluctuate enough to make a rotation-invariant
potential (this is what 
happens in Hamiltonian-strong-coupling perturbation theory 
\cite{RichardsdonPearsonShigemitsu}).

\section{Longitudinal confinement}
\setcounter{equation}{0}
\renewcommand{\theequation}{5.\arabic{equation}}

Next we will show how longitudinal confinement takes place. Again,
the reasoning is essentially that of  Reference \cite{PhysRevD71}.

Let us next consider a quark and antiquark separated in the $3$-direction
only, with coordinates $u^{\perp},u^{3}$ and $u^{\perp}, v^{3}$, respectively. If
$g_{0}^{\prime}=0$, there is just a constant potential between the sources, since longitudinal
electric flux costs no energy. If we assume $g_{0}^{\prime}\neq0$ and
(\ref{relative-scales}) instead, then
electric flux does cost some energy. Furthermore this flux will be concentrated along
the line $u^{\perp}$. The reason for this concentration is that if any flux
leaves this line, residual gauge invariance (\ref{relative-scales}) implies that
one of the four bands 
$(u^{\perp},1)$, $(u^{\perp},2)$, $(u^{\perp}-{\hat 1}a,1)$, $(u^{\perp}-{\hat 2}a,2)$
cannot be in a singlet state. Hence at least one of these four bands is excited with
an energy at least the sigma-model gap $m_{1}$. The mass
gap tends to prevent flux spreading transversely. 

We can now estimate the longitudinal
string tension. It is simply the string tension
of a Yang-Mills theory in one space and one time dimension,
with coupling $g_{0}^{\prime}/a$: 
\beq
\sigma_{\rm L}=
\frac{(g_{0}^{\prime})^{2}}{a^{2}}C_{N}\;, \label{longitudinalsigma}
\eeq
where
$C_{N}$ is the Casimir of 
${\rm SU}(N)$. Notice that the physical mechanism of longitudinal 
confinement is a dual Meissner effect, though no assumptions of
magnetic condensation have been made. 

Note that if we simply took $\lambda=0$ \cite{Verlinde-squared}, we would
have transverse confinement, but no longitudinal confinement. The term
$H^{\prime}$ is essential for the longitudinal string tension.

Corrections of higher order in $(g_{0}^{\prime})$ to (\ref{longitudinalsigma}) come from virtual
pairs of FZ particles. The analogous calculation in $(2+1)$ dimensions has already
been done \cite{PhysRevD74}.

\section{The effective Hamiltonian as an eikonal approximation for QCD}
\setcounter{equation}{0}
\renewcommand{\theequation}{6.\arabic{equation}}

We now argue that the effective Hamiltonian is an eikonal approximation
for QCD. A
conventional approach to this approximation 
\cite{KabatJackiwOrtiz} is to take the fields from
one incoming particle, boost them to the lab frame, and consider the wave function
of the second particle in the presence of those fields (afterwards one can improve the
result by iteratively imposing unitarity). We point out in this section
that the rescaling of the coupling constant in (\ref{action}), (\ref{ContHamiltonian}) and (\ref{hamilt1})
produces ``boosted" electric fields. 

The value of $g_{0}^{\prime}$ must be chosen so that the ratio of the longitudinal string tension to the transverse string tension is small. This ratio, from (\ref{transversesigma}) and 
(\ref{longitudinalsigma}) is
\beq
\frac{\sigma_{\rm L}}{\sigma_{\perp}}=\frac{(g_{0}^{\prime})^{2}}{m_{1}a}
=\frac{\lambda^{2}g_{0}^{2}}{m_{1}a}
\nonumber
\eeq
We can turn this expression around to
find
\beq
\lambda^{2}=\frac{m_{1}a}{g_{0}^{2}} \frac{\sigma_{\rm L}}{\sigma_{\perp}}\;.
\label{firstexpression}
\eeq
Suppose now we consider two incoming hadrons, traveling in the $3$-direction, 
with velocity $\pm v$. If the electric and magnetic
fields of a hadron in its rest frame are
${\mathcal E}^{\prime}_{k}$ and ${\mathcal B}^{\prime}_{k}$, respectively, then in the lab frame
\beq
{\mathcal E}_{1}=\frac{{\mathcal E}^{\prime}_{1}\pm v {\mathcal B}^{\prime}_{2}}{\sqrt{1-v^{2}}},
\;
{\mathcal E}_{2}=\frac{{\mathcal E}^{\prime}_{2}\mp v {\mathcal B}^{\prime}_{1}}{\sqrt{1-v^{2}}},
\;
{\mathcal E}_{3}={\mathcal E}^{\prime}_{3}\;. 
\label{FieldTrans}
\eeq
If we assume the magnetic fields inside the hadron are random, then the average
of the ratio of the longitudinal electric field to the transverse electric field
is $\sqrt{1-v^{2}}$. We substitute this for $\sigma_{L}/\sigma_{\perp}$ in 
(\ref{firstexpression}) to obtain
\beq
\lambda^{2}=\frac{m_{1}a}{g_{0}^{2}} \sqrt{1-v^{2}}\;.
\label{expressionForLambda}
\eeq
Note that (\ref{expressionForLambda}) means that large velocity $v$ implies
small $\lambda$. We have thereby interpreted the rescaling of the coupling
constants in (\ref{action}), (\ref{ContHamiltonian}) and (\ref{hamilt1})
as the result of hadrons moving at high velocities. To make
such an interpretation sensible, however, the velocity in the transformed
coordinates (that is, the coordinates defined after the rescaling) should be
small. Otherwise, the expression (\ref{firstexpression}) cannot be used. This
means that to consider large-$s$ processes,
$\lambda$ should be taken small, but velocities in our new coordinates 
should be nonrelativistic.

Our justification of the rescaling as an eikonal approximation is rather heuristic. It
seems strange, because it contradicts the fact that under a simple rescaling,
the longitudinal component of velocity does not
change. We are arguing that there really is an anomalous transformation
of the velocity. Before rescaling the velocity is close to zero, but
afterwards, it is given by (\ref{expressionForLambda}). It
seems worth making the argument more rigorous. This could conceivably be done
with anisotropic renormalization-group methods \cite{Aref'eva}, 
\cite{Aref'evaVolovich}, \cite{PhysRevD75-2}, \cite{CompositeStrings}.

\section{Hadronic states and diffractive scattering}
\setcounter{equation}{0}
\renewcommand{\theequation}{7.\arabic{equation}}

The structure of hadrons for our effective action is rather similar to that of the strong-coupling
picture \cite{Kogut-Susskind}, despite the fact that only one coupling $g_{0}^{\prime\prime}$,
is strong, all others being weak (the same is true in the $(2+1)$-dimensional case where
all couplings are weak). Hadrons are built out of strings connecting quarks. Strings terminate
at quarks and $N$ of them can meet in junctions. Thus we have a string-parton picture,
in which the partons are quarks and FZ particles.

\begin{figure}[ht]

\begin{picture}(150,200)(30,0)

\put(5,5){\vector(1,0){80}}
\put(88,3){$x^{1}$}

\put(5,5){\vector(0,1){80}}
\put(2,88){$x^{2}$}

\linethickness{0.05mm}
\multiput(10,20)(0,20){9}{\line(1,0){180}}

\multiput(20,10)(20,0){9}{\line(0,1){180}}

\put(65,37){$q$}

\put(60,40){\circle{6}}
\put(60,40){\circle*{2}}

\put(60,50){\circle{11}}
\put(60,70){\circle{11}}

\put(130,120){\circle{11}}
\put(110,120){\circle{11}}
\put(100,110){\circle{11}}
\put(60,40){\circle*{3}}
\put(60,50){\circle*{3}}
\put(60,70){\circle*{3}}
\put(80,90){\circle*{3}}
\put(80,90){\circle{11}}

\put(70,80){\circle{11}}
\put(66,76){\large B}

\put(90,100){\circle*{3}}

\put(90,100){\circle{11}}

\put(110,120){\circle*{3}}
\put(130,120){\circle*{3}}
\put(100,110){\circle*{3}}

\put(100,150){\circle*{3}}
\put(100,150){\circle{11}}

\put(100,160){\circle{6}}
\put(100,160){\circle*{2}}
\put(105,160){$q$}

\put(180,120){\circle{6}}
\put(180,120){\circle*{2}}

\put(150,120){\circle*{3}}
\put(170,120){\circle*{3}}

\put(150,120){\circle{11}}
\put(170,120){\circle{11}}

\put(185, 120){$q$}

\put(100,130){\circle{11}}
\put(100,130){\circle*{3}}

\put(117,158){$\otimes$}
\put(125,158){$q$}

\put(120,150){\circle{11}}
%\put(120,150){\circle{10}}
%\put(120,150){\circle{9}}
\put(116,146.5){\large X}

\put(120,130){\circle{11}}
%\put(120,130){\circle{10}}
%\put(120,130){\circle{9}}
\put(116,126){\large X}

\put(116,116){\large A}

\put(120,110){\circle{11}}
%\put(120,110){\circle{10}}
%\put(120,110){\circle{9}}
\put(116,106){\large X}

\put(120,90){\circle{11}}
\put(116,86){\large X}

\put(110,80){\circle{11}}
\put(106,76){\large X}

\put(90,79.5){\circle{11}}
\put(86,76){\large X}

%\put(70,80){\circle{11}}
%\put(66,76){\large X}

\put(50,80){\circle{11}}
\put(46,76){\large X}

\put(40,70){\circle{11}}
\put(36,66){\large X}

\put(30,60){\circle{11}}
\put(26,56){\large X}

\put(20,70){\circle{11}}
\put(16,66){\large X}

\put(20,90){\circle{11}}
\put(16,86){\large X}

\put(20,110){\circle{11}}
\put(16,106){\large X}

\put(17,118){$\otimes$}

\put(25, 117.75){$q$}

\put(130,80){\circle{11}}
\put(126,76){\large X}

\put(150,80){\circle{11}}
\put(146,76){\large X}

\put(157,78){$\otimes$}

\put(165,78){$q$}

\linethickness{0.4mm}

\put(20,118){\line(0,-1){58}}
\put(20,60){\line(1,0){20}}
\put(40,60){\line(0,1){20}}
\put(40,80){\line(1,0){27}}
\put(74,80){\line(1,0){83}}
\put(60,80){\line(0,-1){37}}

\put(80,80){\line(0,1){20}}
\put(80,100){\line(1,0){20}}
\put(100,100){\line(0,1){57}}
\put(100,120){\line(1,0){16}}
\put(123,120){\line(1,0){54}}

\put(120,80){\line(0,1){36}}

\put(120,123){\line(0,1){37}}

\end{picture}
\caption{Longitudinally approaching baryons. Large circles
represent FZ particles and small circles represent quarks.
The symbol $\odot$ means that the constituent
particle (FZ particle or quark) is coming out of the page and the symbol $\otimes$ means
that it is going into the page. Strings of FZ particles can intersect at lines of fixed $x^{\perp}$
(A) or overlap at transverse bands (B). }

\end{figure}
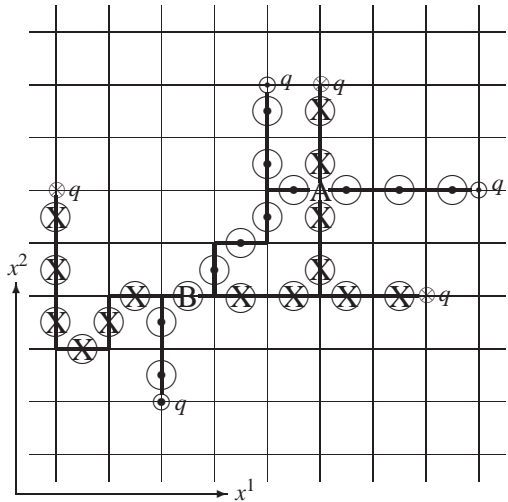

The transverse projection of a state of two baryons approaching each other longitudinally
is shown in Fig. 5. The hadrons can collide at coincident lines or bands. Let
us suppose we neglect
$H^{\prime}+H^{\prime\prime}$. In the case
of overlapping sites (but without overlapping at
adjacent links), longitudinal flux exchange (LFE) at lines of fixed
$x^{\perp}$ can happen, but nothing else
occurs. The LFE amplitude can be thought of as due to resonance between different flux
arrangements and is of order $1/N$, as we explain in the next paragraph.

The LFE process is analogous to the case of meson
scattering in $(1+1)$ dimensions. We can think of the two FZ particles 
adjacent to the line $x^{\perp}$ (in one hadron) as being similar to
two sources of fundamental color in that line (recall that
FZ particles are dipoles, and
each of them has one fundamental source adjacent to the line). Thus
LFE is essentially similar to flux exchange in $(1+1)$-dimensional
QCD. This non-planar process is of order
$1/N$ \cite{'tHooft}. An example of LFE is shown in Fig. 6.

We note that world sheet of longitudinal electric flux lines during LFE has
the same topology as that of Pomeron exchange in string models.

In the case of two hadrons intersecting the same band, {\em i.e.} with
colliding FZ particles, the
S-matrix of the principal-chiral sigma model comes into play. We note that this $(1+1)$-dimensional
S-matrix
is unity in the large-$N$ limit, $N\rightarrow \infty$, $g_{0}^{2}N$ fixed. Thus 
in this limit, no
interaction occurs. In the $1/N$ expansion of this S-matrix, however, LFE occurs 
\cite{abda-wieg}.

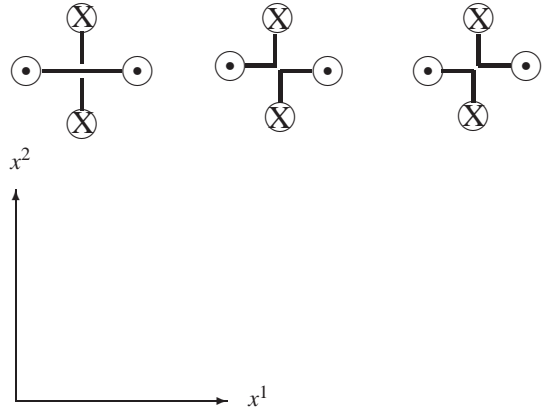
\begin{figure}[t]

\begin{picture}(150,200)(30,20)

\put(0,25){\vector(1,0){80}}
\put(88,23){$x^{1}$}

\put(0,25){\vector(0,1){80}}
\put(-2,113){$x^{2}$}

\linethickness{0.5mm}

\put(10,150){\line(1,0){30}}
\put(4,150){\circle{11}}
\put(4,150){\circle*{3}}
\put(46,150){\circle{11}}
\put(46,150){\circle*{3}}

\put(25,135){\line(0,1){12}}
\put(25,153){\line(0,1){12}}

\put(25,130){\circle{11}}
\put(21,126.5){\large X}

\put(25,170){\circle{11}}
\put(21,166.5){\large X}

\put(87,152){\line(1,0){11.5}}
\put(100,150){\line(1,0){12}}

\put(100,138){\line(0,1){12}}
\put(98,152){\line(0,1){12}}

\put(81,152){\circle{11}}
\put(81,152){\circle*{3}}
\put(118,150){\circle{11}}
\put(118,150){\circle*{3}}
\put(100,132.5){\circle{11}}
\put(95.5,128.5){\large X}
\put(99,170){\circle{11}}
\put(95,166){\large X}

\put(173,138){\line(0,1){12}}

\put(175,152){\line(0,1){12}}

\put(161,150){\line(1,0){12}}

\put(175,152){\line(1,0){12}}

\put(156,150){\circle{11}}
\put(156,150){\circle*{3}}
\put(193,152){\circle{11}}
\put(193,152){\circle*{3}}
\put(173,133){\circle{11}}
\put(169,129){\large X}
\put(175,170){\circle{11}}
\put(171,166.){\large X}

\end{picture}

\caption{An example of
longitudinal flux exchange (LFE) is resonance between the configurations
shown here, resembling (A) in Figure 5. Longitudinal 
flux lines are present, in general, in the second
and third diagrams. The particles shown can be either quarks or FZ particles.}

\end{figure}

\section{Conclusions}
\setcounter{equation}{0}
\renewcommand{\theequation}{8.\arabic{equation}}

In this paper, we have considered the anisotropic effective
theory of Reference \cite{Verlinde-squared}, which is a longitudinally-rescaled Yang-Mills
theory, with rescaling parameter $\lambda$. We have shown that
this theory confines for sufficiently small $\lambda$ and a natural
string-parton picture emerges. The partons are valence
quarks and the soliton-like FZ particles of the principal-chiral
non-linear sigma model. The interpretation of this effective action as
an eikonal approximation is rather subtle, since electric fields are much
stronger in the  transverse than the longitudinal direction, even for slow-moving
hadrons. We have argued that this interpretation is correct, provided colliding
hadrons are moving anomalously slowly 
in the rescaled coordinates. An important
process in hadronic scattering
in the forward direction is the exchange of longitudinal flux, even
if FZ particles collide.

There are essentially two areas which come to mind for further
investigation.

We would like to improve our understanding
of confinement for the anisotropic gauge theory
considered here. This is not a strong-coupling gauge theory, but 
rather a hybrid gauge theory where one coupling, namely $g_{0}^{\prime\prime}$, 
is strong and the rest 
are weak. A demonstration that confinement still holds if 
$g_{0}^{\prime\prime}$ is
weak  would be real progress on the QCD confinement problem. We have
been investigating expansions in $(g_{0}^{\prime\prime})^{-2}$ using
field form factors of the sigma model
\cite{BalogWeisz}, but as yet have no simple argument
that confinement holds for small $g_{0}^{\prime\prime}$. In 
any case, it seems possible to generalize our calculations
of the corrections to the longitudinal \cite{PhysRevD74} and
string tensions \cite{CompositeStrings},  
and to the mass spectrum \cite{PhysRevD75-2}
to $(3+1)$ dimensions, with $g_{0}^{\prime\prime}$ large, but
not infinite.

The other problem of importance is the forward scattering amplitude
of hadron-hadron scattering. We hope that its solution
will yield a quantitative understanding of the Pomeron. The solution
will require some 
control of LFE processes. We believe that this problem is tractable. We
have argued in Section 6 that in the
new coordinates, momenta should be taken as small
as possible; in this way the electric fields are longitudinally boosted as they
should be. Thus, LFE processes can be studied
in a nonrelativistic context. We hope to make progress
on this problem soon. Some assumptions may need to be
made for the distributions of partons within a hadron. A 
good starting point should be the solution of
hadron-hadron scattering in $(2+1)$ dimensions, where
the parton distributions are simpler.

\begin{acknowledgments}

We thank Gregory Korchemsky for a brief, but
very informative discussion, Jamal 
Jalilian-Marian for inspiration and Poul Henrik 
Damgaard for encouragement.

This research was supported in
part by the National Science Foundation, under Grant No. PHY0653435
and by a grant from the PSC-CUNY.

\end{acknowledgments}


\begin{thebibliography}{xx}
\bibitem{Lipatov} L.N. Lipatov, Nucl. Phys. {\bf B365} (1991) 614.
\bibitem{McLerranVenugopalan} L. McLerran and R. Venugopalan, Phys. Rev.
{\bf D49} (1994) 2233; {\bf D49} (1994) 3352; 
{\bf D50} (1994) 2225; {\bf D59} (1999) 094002. 
\bibitem{Kovchegov} Y. Kovchegov, Phys. Rev. {\bf D54} (1996) 5463; 
{\bf D55} (1997) 5445. 
\bibitem{Jalilian-Marian-et-al} J. Jalilian-Marian, A. Kovner, L. McLerran, H. Weigert, 
Phys. Rev. {\bf D55} (1997) 5414; 
J. Jalilian-Marian, A. Kovner, A. Leonidov and H. Weigert, Nucl. Phys. {\bf B504}
(1997) 415; Phys. Rev. {\bf D59} (1999) 014014; {\bf D59} (1999) 034007; J. 
Jalilian-Marian, A. 
Kovner and H. Weigert, Phys. Rev. {\bf D59} (1999) 014015;  J. 
Jalilian-Marian and X.N. Wang, Phys. Rev. {\bf D60} (1999) 054016; A. 
Kovner and J.G. Milhano, Phys.Rev. {\bf D61} (2000) 014012.
\bibitem{Jalilian-Marian-et-al-2} J. Jalilian-Marian, S. Jeon and
R. Venugopalan, Phys.Rev. {\bf D63} (2001) 036004.
\bibitem{Verlinde-squared}  H. Verlinde and E. Verlinde, Princeton University
Preprint {\bf PUPT-1319}, {\bf hep-th/9302104} (1993).
\bibitem{KirschLipSzyman} R. Kirschner, L.N. Lipatov and L. Szymanowski,
{\bf B425} (1994) 597; Phys. Rev. {\bf D51} (1995) 838.
\bibitem{Hatta-etal} Y. Hatta, E. Iancu, L. McLerran, A. Stasto and
D.N. Triantafyllopoulos, Nucl. Phys. {\bf A764} (2006) 423. 
\bibitem{Kogut-Susskind} J.B. Kogut and L. Susskind, Phys. Rev. {\bf D11}  (1975) 395.
\bibitem{PhysRevD71} P. Orland, Phys. Rev. {\bf D71} (2005) 054503. 
\bibitem{PhysRevD74} P. Orland, Phys. Rev. {\bf D74} (2006) 085001.
\bibitem{PhysRevD75-1} P. Orland, Phys. Rev. {\bf D75} (2007) 025001.
\bibitem{PhysRevD75-2} P. Orland, Phys. Rev. {\bf D75} (2007) 101702(R).
\bibitem{CompositeStrings} P. Orland, Baruch Preprint {\bf BCCUNY-HEP/07-05},
{\bf 0710.3733 [hep-th]} (2007), to be published in Physical Review {\bf D}.
\bibitem{BardeenPearsonRabinovici} W.A. Bardeen and R.B. Pearson, Phys. Rev.
{\bf D14} (1976) 547; W.A. Bardeen, R.B. Pearson and E. Rabinovici, Phys. Rev.
{\bf D21} (1980) 1037.
\bibitem{Aref'eva} I.Ya. Aref'eva, Phys. Lett. {\bf B325} (1994) 171; 
Phys. Lett. {\bf B328} (1994) 411.
\bibitem{Aref'evaVolovich} I.Ya. Aref'eva and I.V. Volovich, Steklov Math. Inst. Preprint
{\bf SMI-15-94}, {\bf hep-th/9412155} (1994).
\bibitem{Patel} A.D. Patel, Nucl. Phys. Proc. Suppl. {\bf B94}
(2001) 260; and R. Ratabole, Nucl. Phys. Proc. Suppl.{\bf B129} (2004) 889.
\bibitem{abda-wieg}  E. Abdalla, M.C.B. Abdalla and A. Lima-Santos,
Phys. Lett. {\bf B140} (1984) 71; P.B. Wiegmann; Phys. Lett. {\bf B142} (1984) 173;
A.M. Polyakov and P.B.  Wiegmann, Phys. Lett. {\bf B131} (1983) 121;  P.B. 
Wiegmann, Phys. Lett. {\bf B141} (1984) 217.
\bibitem{KarowskiWeisz} M. Karowski and P. Weisz, Nucl. Phys. {\bf B139} (1978) 455.
\bibitem{LipatovKorchemsky} L.N. Lipatov,
{\bf hep-th/9311037} (1993), L.D. Faddeev and G.P. Korchemsky, Phys. Lett.
{\bf B343} (1995) 311.
\bibitem{BalitskiFadKurLip} V. Fadin, E. Kuraev and L.N. Lipatov, Sov. Phys. J.E.T.P {\bf 44} 
(1976) 443; Ya. Balitski and L.N. Lipatov, Sov. Nucl. Phys. {\bf 28} (1978) 822.
%\bibitem{ColorGlass} L.D. McLerran, Lectures given at the
%40$^{\rm th}$ Schlamding Winter School: {\bf Dense Matter}, March 3-10, 2001, {\bf
%hep-ph/0104285} (2001); E. Iancu, A. Leonidov and L.D. McLerran, Lectures given
%at the Cargese Summer School, {\bf QCD Perspectives on Hot and Dense Matter}, Cargese, 
%France,August 6-18, 2001, {\bf hep-ph/0202270} (2001); E. Iancu, R. Venugopalan,
%{\bf Quark-Gluon Plasma 3}, eds R.C. Hwa and X.N. Wang, World Scientific, Singapore,
%{\bf hep-ph/0303204} (2003).
\bibitem{BalogWeisz} J. Balog and P. Weisz, Nucl. Phys. {\bf B778} (2007)
259.
\bibitem{Mandelstam} S. Mandelstam, Phys. Reports {\bf 23} (1976) 245; Bull. Amer. Phys. Soc.
{\bf 22} (1977) 541;  Phys. Rev. {\bf D19} (1979) 2391.
\bibitem{RichardsdonPearsonShigemitsu}
J. B. Kogut, D. K. Sinclair,
R. B. Pearson, J. L. Richardson and J. Shigemitsu,
Phys. Rev. {\bf D23} (1981) 2945.
\bibitem{KabatJackiwOrtiz} G. 't Hooft, Phys. Lett. {\bf B198} (1987) 61; Nucl. Phys. {\bf B304}
(1988) 867; R. Jackiw, D. Kabat and M. Ortiz, Phys. Lett. {\bf B277} (1992) 148.
\bibitem{'tHooft} G. 't~Hooft, Nucl. Phys. {\bf B75} (1974) 461.
\end{thebibliography}
\end{document}